\documentclass[reprint,amsmath,amssymb,mathtools,aps,pra]{revtex4-2}

\usepackage{graphicx}
\usepackage{subcaption}
\usepackage{dcolumn}
\usepackage{bm}
\usepackage{lipsum}
\usepackage{xcolor}
\usepackage{multirow}

\begin{document}


\title{Ultracold long-range van der Waals Rydberg trimers}

\author{Mateo Londoño}
 \email{mateo.londoo@stonybrook.edu}
\author{Jesús Pérez-Ríos}%
\affiliation{%
Department of Physics and Astronomy, Stony Brook University, Stony Brook, NY 11794, USA}%
\author{Vanessa C. Olaya-Agudelo}%
\author{Felipe Herrera}
 
\affiliation{Department of Physics, Universidad de Santiago de Chile, Santiago 9170124, Chile}%

\date{\today}

\begin{abstract}
Rydberg molecules, Rydberg-atom or Rydberg-molecule, are an essential ingredient of cold molecular sciences. However, due to the richness of Rydberg-neutral interactions, new kinds of Rydberg molecules and binding mechanisms are still to be discovered.  In this work, we predict the existence of ultra-long-range van der Waals trimers in dilute atom-gas mixtures. These are bound states of a Rydberg atom and a diatomic polar molecule mediated by the long-range van der Waals interaction. This new binding mechanism gives rise to trimers with sizes between 5-500~nm and binding energies between 2 MHz and 0.2 kHz depending on the atomic principal quantum number $n$ and orbital angular momentum $L$. We show that these molecules can be produced via two-photon photoassociation, with rates on the order of (10$^{-13}$ - 10$^{-11}$) cm$^{3}$s$^{-1}$ for temperatures in the range of (0.5 $\mu K$ - 10$\mu K$), and discuss the feasibility of observing trimer resonances.

\end{abstract}

--\maketitle


\section{\label{sec:introcduction} Introduction}

Rydberg atoms play a pivotal role in several areas of atomic, molecular, and optical physics \cite{Gallagher2005rydberg}, such as the implementation of novel quantum information protocols~\cite{Saffman2010}, quantum simulation of many-body Hamiltonians~\cite{Many_body_Rydberg}, the study of impurity physics, non-linear quantum optics, and ultracold chemistry~\cite{Michael2016b,JPRBook,Karman2024}. Many of these applications rely on controlling the exaggerated properties of Rydberg states in order to tailor the resulting interatomic forces. For instance, at high densities, Rydberg excitations present a rather unique lineshape as a consequence of the Rydberg electron-perturber atom interaction~\cite{Liebisch2016,Trevor}, leading to the formation of ultralong-range Rydberg molecules~\cite{Greene2000,Bendkowsky2009,Hamilton2002,Khuskivadze2002,Gaj2014,Trilobite,Butterlfy,ULR_1,ULR_2,ULR_3,Eiles2016}, and possible polaron effects, considering the Rydberg atom as an impurity in a dense atomic reservoir.~\cite{Rydberg_Polaron,Rydberg_Polaron2,Eiles2024} Another key ingredient of atomic, molecular, and optical physics is cold molecules, an essential platform for exploring fundamental quantum phenomena. For instance, controlling interactions enables precise studies of quantum chemistry~\cite{Karman2024}, novel quantum phases~\cite{Bigagli2023}, and quantum information processing~\cite{Cornish2024}. Their long-range interactions and tunable states make them ideal for simulating complex quantum systems and advancing quantum technology applications. Cold molecules are also suitable for creating hybrid-quantum systems to study atomic or charged impurity dynamics~\cite{Langen2024, Wenzel2018}.

\begin{figure*}[ht]
\centering
\includegraphics[width=0.9\textwidth]{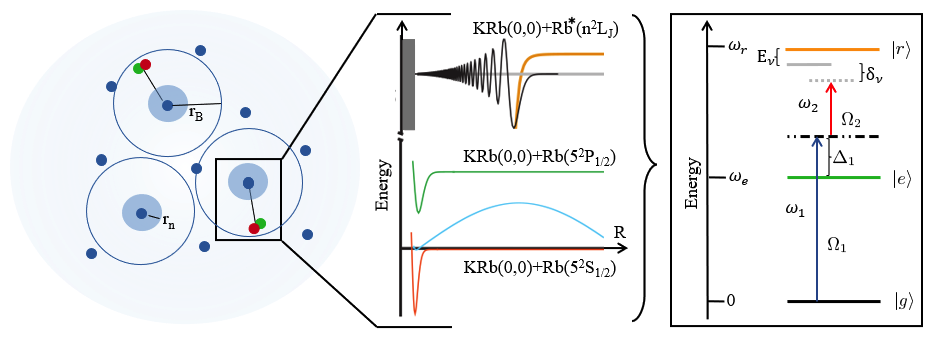}  
\caption{{\bf Formation of ultra-long range Rydberg trimers}. Left: Scheme of the atomic and molecular diluted gases, where not more than a single molecule can be found within the Rydberg blockade radius $r_B$, but outside the Rydberg electron orbit radius $r_n$. Center: 
Example of interacting channels and potentials involved in the PA process for a given atom-molecule pair (Rb + KRb). The relevant scattering and bound-state wavefunctions are shown. The process efficiency depends on the overlap between the initial scattering state [KRb(0,0)+Rb($5^2S_{1/2}$)] and the final near-threshold vibrational state of the electronically excited trimer [KRb(0,0)+Rb$^*$($n^2L_j$)]. The substantial difference of van der Waals lengths between the initial and final states is highlighted. Right: Two-photon scheme for exciting alkali-metal atoms into a target Rydberg state, where atomic states $|g\rangle, |e\rangle, |r\rangle$ are connected through the lasers $\omega_1$ and $\omega_2$, giving the two-photon detuning $\delta_\nu=\omega_1+\omega_2-(\omega_r-E_\nu))$ from the trimer bound level $E_\nu$. Large intermediate detunings $\Delta_1$ from the excited atomic level prevent gas heating via light scattering.}
\label{fig:scheme}
\end{figure*}

In combination with cold molecules, Rydberg atoms offer a unique opportunity for studying exotic Rydberg molecules. For instance, a Rydberg excitation in a dense dipolar gas can lead to the formation of an ultra-long-range Rydberg molecule bound state via elastic collisions of the Rydberg electron with the molecule~\cite{Ferez2017,Ferez2020}. When the Rydberg excitation is induced in a molecule, it gives rise to the formation of Rydberg bimolecules, an ultra-long range Rydberg molecule in which a molecular ionic core is bound to a molecule (inside the Rydberg orbit) via the scattering of the Rydberg electron off the perturbing molecule. These exotic molecules show binding energies in the GHz regime and kilo-Debye permanent dipole moments~\cite{Ferez2021}. A similar scenario has been proposed as a technigue for sympathetic cooling of molecules~\cite{Sympahteticcooling}.

In this work, we predict the existence of another type of formed long-range Rydberg trimer molecule when a Rydberg atom binds to a diatomic polar molecule via van der Waals interaction, in a regime where the molecular density is much lower than the density of Rydberg atoms so that at most one molecule resides within the Rydberg blockade radius $r_B$, as displayed in Fig.~\ref{fig:scheme}. The long-range van der Waals Rydberg trimer is formed by a two-photo photoassociation scheme from the ground state of the atom and the molecule. In this case, in analogy to the lineshape of Rydberg excitations in atomic gases \cite{Gallagher2005rydberg}, the two-photon photoassociation lineshape spectra will show traces of the existence of long-range van der Waals Rydberg trimers.

The Van der Waals interaction, responsible of the bonding in these new trimers, dominates when the Rydberg-molecule distance is larger than the Le-Roy Radius. In this sense, the predicted trimers are formed in a density regime that is complementary to previous reported ultralong-range Rydberg molecules, where the oscillatory Born-Oppenheimer potentials are formed below the Le-Roy radius \cite{Mellado2024}. From this we expect to have exotic Rygdberg trimers either the molecule is inside or outside the Rydberg's electron orbital, with a particular transition in the bonding mechanism at the limit of the two regions.

\section{\label{Theoretical Framework} Theoretical Framework}
\subsection{Two photon photoassociation rates}
Let us consider a mixed thermal gas consisting of atoms, X, and molecules, AB, in their ground vibrational state. A two-photon excitation resonant with a Rydberg state of the atom, X$^*$, induces the following light-assisted chemical reaction, known as two-photon photoassociation (PA)~\cite{Jones}:
\begin{equation}
\label{eq1}
\text{X} + \text{AB} + 2\gamma \rightarrow \text{X}^*-AB,
\end{equation}
 where X$^*$-AB represents an excited trimer resonant with an excited electronic state that correlates with the atomic Rydberg state of the atom. These excited states are the long-range van der Waals trimers, and their existence is contingent to the nature of the long-range van der Waals X$^*$ + AB interaction and the reaction rate of Eq.~(\ref{eq1}).

We consider our atom as $^{85}$Rb($5^2S_{1/2}$) and the molecule, either LiRb or KRb molecules in the ground rovibrational state ($^1\Sigma^+,v=0, J=0$) since both of these molecules are available in the cold and ultracold regimes~\cite{Dutta,Blasing2016,Ni2008}. The Rydberg state of the atom is denoted as $n^2L_j$, where $L$ is the atomic orbital angular momentum and $j$ is the total electronic angular momentum. The initial state of the PA process is a scattering  atom+molecule state denoted $|\Psi_l(E_{\text{kin}}) \rangle$, with collision energy $E_{\text{kin}}$ and partial wave $l$, in the ground atom-molecule channel $|5^2S_{1/2}\rangle |X^1\Sigma^{+} \rangle$. Here, we assume an $s$-wave collision between ($l = 0$). The final state of the process $|\Psi_{\nu}\rangle$ corresponds to the $\nu$-th bound state of the trimer Rydberg-molecule potential that correlates with the $|n^2L_{j}\rangle |X^1\Sigma^{+} \rangle$ asymptote. Here, we follow the usual labeling for vibrational states as $\nu = -1, -2, -3, \dots$, where $\nu = -1$ denotes the shallowest bound state.

\begin{figure*}[ht]
\centering
\includegraphics[width=1\textwidth]{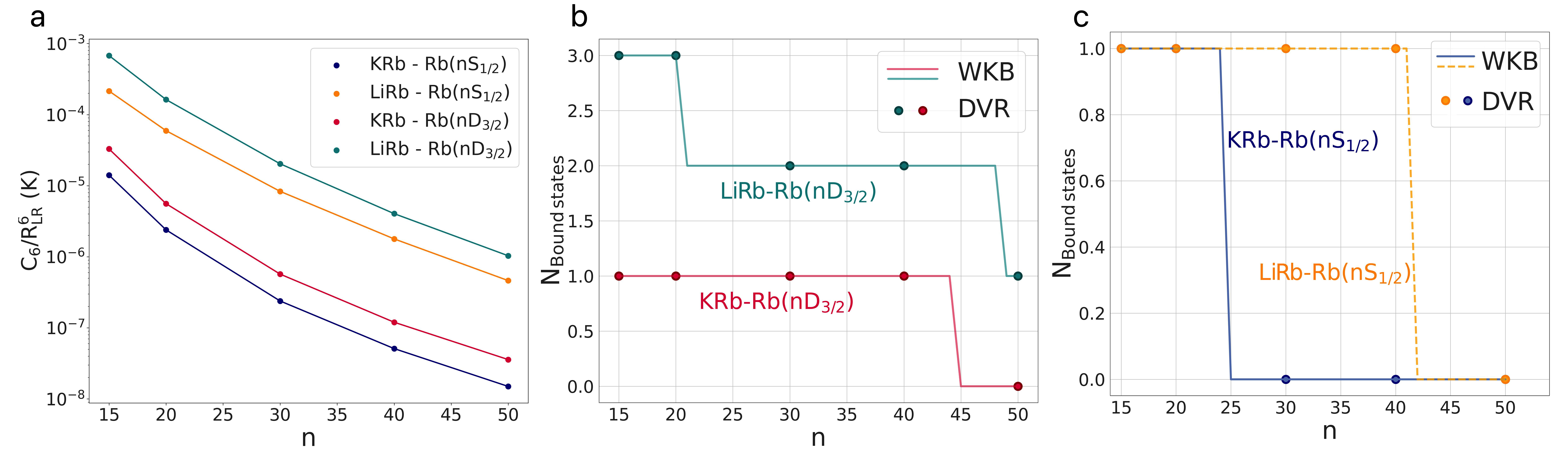}  
\caption{{\bf Characterization of the long-range Rydberg-molecule energy landscape.} Panel (a) shows the dissociation energy of the trimer, it means, the Van der Waals potential evaluated at the minimum distance R$_{LR}$. Panels (b) and (c) presents the number of bound states of each Rydberg-molecule potential, for the $D$ and $S$ states respectively. Both panels show the results as a function of the principal quantum number and the species considered.}
\label{fig:depths_bounds}
\end{figure*}

For distances beyond the Le Roy radius of the Rydberg atom $R_{\rm LR}$ \cite{LeRoy}, the Rydberg-molecule interaction is given by the van der Waals term $V(r)=-C_6/r^6$, where $r$ is the Rydberg-molecule distance. In many cases, this interaction has been shown to be attractive~\cite{Olaya}, so it can support bound states depending on the magnitude $C_{6}$. The characterization of the Rydberg-molecular interaction is shown in Figure~\ref{fig:depths_bounds}a, displaying the Rydberg-molecule potential depths as a function of the atomic main quantum number $n$ for all the species being considered. It can be seen that existing bound states energies range between 0.2~kHz to 2~MHz. We computed the number of bound states for each species using the  Wentzel–Kramers–Brillouin (WKB) approximation, and the results agree with exact numerical calculations as shown in panels (b) and (c) of Figure~\ref{fig:depths_bounds}. Highly excited $n$-states have a larger Le Roy radius, which reduces the effective range of the attractive van der Waals force. As a result, the potential well becomes shallower, limiting the number of bound states. The $C_6$ coefficient, which depends on the molecular structure, is an order of magnitude larger for LiRb than for KRb, explaining the observed difference in the number of bound states~\footnote{ For example, for Rb$^*$(15D$_{1/2}$), the van der Waals coefficients are $C_{6}$(LiRb) = -2.778$\times$10$^{8}$ a.u, and $C_{6}$(KRb) = -1.351$\times$10$^{7}$ a.u. The Le Roy radius is the same in both cases, approximately 712 a$_{0}$.}

We compute PA rates for $15\leq n \leq 50$ Rydberg levels in the $^2S_{1/2}$ and $^2D_{3/2}$ atomic channels, but the formalism can be equally applied to other atomic and molecular states, provided the $C_6$ coefficients in the initial and final states are known. From resonance scattering theory~\cite{Jones, Bohn}, the PA rate constant can be written as
\begin{equation}
\label{eq2}
K_{PA} (E_{\text{kin}}) =  v_{\text{rel}} \frac{\pi}{k^2} \sum_{l=0}^\infty (2l+1) |S_{\nu}(E_{\text{kin}},l)|^2,
\end{equation}

\noindent
where $v_{\text{rel}}= \sqrt{8 k_B T / \pi \mu}$ is the relative velocity of the particles in the entrance channel, $\mu$ is the reduced mass of the Rydberg atom-molecule system, and $k = \sqrt{2\mu k_B T / \hbar^2}$ is the channel wavevector. $k_B$ is the Boltzmann's constant and $T$ is the temperature. The rate constant is determined by the element of $S$-matrix that connects the scattering state and the bound state, given by~\cite{Jones}
\begin{equation}
\label{eq3}
|S_{\nu}(E_{\text{kin}},l)|^2 = \frac{ \gamma\, \Gamma_{\nu}(E_{\text{kin}},l) }{ [ E_{\text{kin}}/\hbar + \delta_\nu ]^2 + [ { \Gamma_{T}}/{2} ]^2}, 
\end{equation}
where $\delta_\nu=\omega_1+\omega_2-(E_a-E_\nu)$ is the two-photon detunning from the final state $|\Psi_{\nu}\rangle$, where $E_{a}$ is the energy of the Rb$^*$($n^2L_{j}$) + Diatom ($X^1\Sigma^{+}, v=0, J=0$) asymptote, and $E_\nu$ is the binding energy of the $\nu$-th near-threshold bound state. We can neglect the possible light-shifts of these resonance frequencies for weak laser dressing fields. The two-photon excitation scheme is further specified in figure \ref{fig:scheme} (right hand side) in terms of the individual Rabi frequencies $\Omega_1$ and $\Omega_2$ of the driving lasers at $\omega_1$ and $\omega_2$, respectively. The derivation of the effective two-level approximation used in Eq.~(\ref{eq3}) is given in the appendix \ref{appendixA}. 

The width of the scattering resonance in Eq.(\ref{eq3}) depends on the overall decay rate of the atom-molecule trimer state, given by $\Gamma_{\text{T}} = \gamma + \Gamma_{\nu}(E_{\text{kin}},l)$, where $\gamma$ is the natural linewidth of the Rydberg state and the stimulated absorption rate is given by 
\begin{equation}
\Gamma_{\nu}(E_{\text{kin}},l) = 2\pi |V_{\nu} (E_{\text{kin}},l)|^2,
\end{equation}
where
\begin{equation}\label{eq:overlap}
V_{\nu} (E_{\text{kin}},l) = \frac{\Omega_{\rm eff}}{2} \langle \Psi_{\nu} | \Psi_l(E_{\text{kin}}) \rangle.
\end{equation} 
Here, $\langle \Psi_{\nu} | \Psi_l(E_{\text{kin}}) \rangle$ represent the Franck-Condon Factor (FCF) between the initial scattering wavefunction and the final bound state wavefunction. For loosely bound vibrational states, it is a good approximation to neglect the radial dependence of the electric transition dipole moment, leading to an effective two-photon Rabi frequency as $\Omega_{\rm eff} = {\Omega_1 \Omega_2}/{2\Delta_1}$ (more details in Ref. ~\cite{Browaeys} and in the appendix \ref{appendixA}).

\begin{figure}[htbp]
\centering
\includegraphics[width=0.41\textwidth]{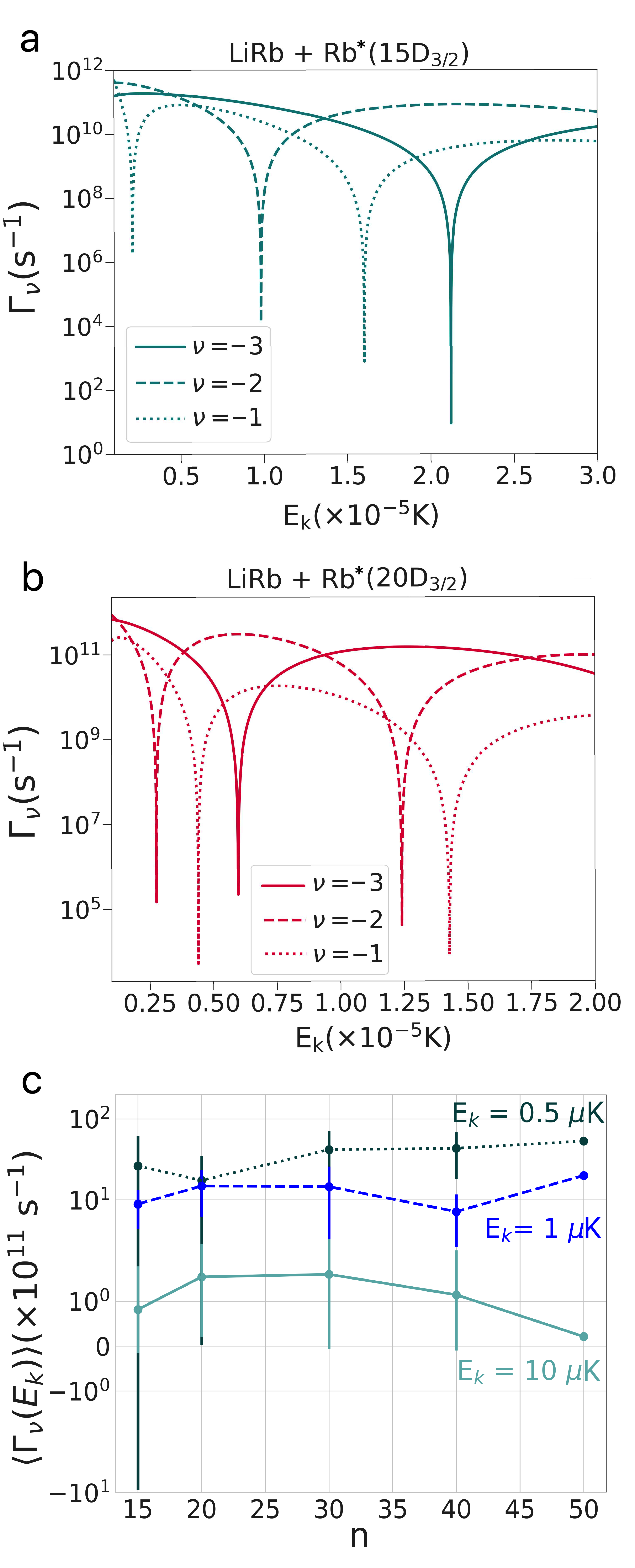}  
\caption{{\bf Stimulated Absorption rates as a function of the kinetic energy of the (LiRb + Rb$^*$) collision pair} Panel (a) presents the stimulated absorption rates for Rb$^*$(15D$_{3/2}$). Rydberg state, equivalently for Panel (b) showing the Rb$^*$(20D$_{3/2}$) state. Panel (c) is the average, over available bound states, absorption rate as a function of the principal quantum number $n$ for LiRb+Rb$^*$($n$D$_{3/2}$) at different collision energies.} 
\label{fig:over}
\end{figure}


The bound state trimer wavefunctions were computed using the mapped Fourier Grid Hamiltonian method \cite{Kokoouline} and the initial scattering wavefunction via the Numerov method. With this information at hand, we compute the stimulated absorption coefficients for LiRb-Rb$^*$(15~D$_{3/2}$) and LiRb-Rb$^*$(20~D$_{3/2}$).

\subsection{Scattering and bound state wavefunctions}
The initial scattering wavefunction was obtained by solving numerically the Schr\"odinguer equation describing the interaction of ground state LiRb(KRb) with ground states Rb. We solve the radial equation with the Numerov method assuming a Lennard-Jones potential between the atom and the molecule. For LiRb interacting with Rb the Lennard-Jones parameters are $C_{6}$ = 7159 a.u, $D_{e}$ = 2036 cm$^{-1}$ and $C_{12}$ = $C_{6}^{2}$/4$D_{e}$, while for KRb interacting with Rb we have  $C_{6}$ = 8798 a.u, $D_{e}$ = 1584 cm$^{-1}$, with the same definition for $C_{12}$. Both set of parameters are taken from Ref.~\cite{Mayle}.

Bound state wavefunctions $| \Psi_{\nu} \rangle$ are calculated using a mapped Fourier grid Hamiltonian method~\cite{Kokoouline} given an Rydberg-molecule effective trimer potential. It was mention that the potential follows the Van der Waals expression V(R) = $C_{6}/R^{6}$ up to R = R$_{c}$ where we impose a repulsive barrier avoiding further penetration. The discussed results were obtain using R$_{c}$ = 1.1R$_{LR}$, where Le Roy radius is estimated from R$_{LR} \sim \sqrt{10}n^{2}a_{0}$. The convergence of the results was tested by varying the position of the barrier between R$_{LR}$ and 1.2 R$_{LR}$. The relative deviation in the energy of threshold states, which are the most affected by the barrier position, ranges from 9$\%$ - 16$\%$, scaling with the value of the Le Roy radius. Additionally, even without knowing the exact barrier position, we can confirm that the energy of the threshold bound state does not exceed the theoretical limit $E_{-1}\approx 39.5\, E_{\rm vdW}$ in the $s$-wave regime as outline in ref. \cite{Gao2000}. Here, $E_{\rm vdW}\equiv \sqrt{2}\hbar^3/\mu^{3/2}C_6^{1/2}$. As a specific example, for the highly excited LiRb+Rb$^{*}$(50D$_{3/2}$) state, the theoretical threshold-state energy is approximately 1.694$\times10^{-2}$MHz, while the reported value we used is 1.062$\times10^{-2}$MHz. The remaining bound states are generally farther from their respective theoretical limits. The obtained number of bound states coincides with analytical estimates from the WKB approximation which yields, for a given $C_{6}$ coefficient, a total number of bound states $N_{max} = \frac{1}{10n^{4}}\sqrt{\frac{\mu C_{6}}{2\pi^{2}}}$ + $\frac{\pi}{4}$. Where the outer turning point is approximated to be at $R = \infty$ and the inner one at $R = R_{\rm{LR}}$, to perform the integral. 

\section{Results}
The results are shown in figure \ref{fig:over} assuming the following Rydberg dressing parameters: $\Delta_1 = 80$ MHz, $\Omega_1 = 10$ MHz and $\Omega_2 = 8$ MHz, corresponding to laser intensities $I_1= 10^{-4}$ W/cm$^2$ and $I_2 = 1.5$ W/cm$^2$ at frequencies $\omega_1$ and $\omega_2$, respectively. The stimulated absorption rates are rather smooth except for certain dips. These are due to the energy-dependent oscillatory nature of the continuum wavefunction that affects its overlap with a given bound wavefunction. The average, with respect to all available bound states, stimulated absorption rate for LiRb+Rb$^*$($n$D$_{3/2}$), on the other hand, shows a weak dependence with the principal quantum number independently of the collision energy. However, at lower collision energies, the rate is systematically larger than that of higher collision energies. Therefore, the observation of the Van der Waals trimer states will benefit from colder temperatures. 

\begin{figure}[htbp]
\centering
\includegraphics[width=0.41\textwidth]{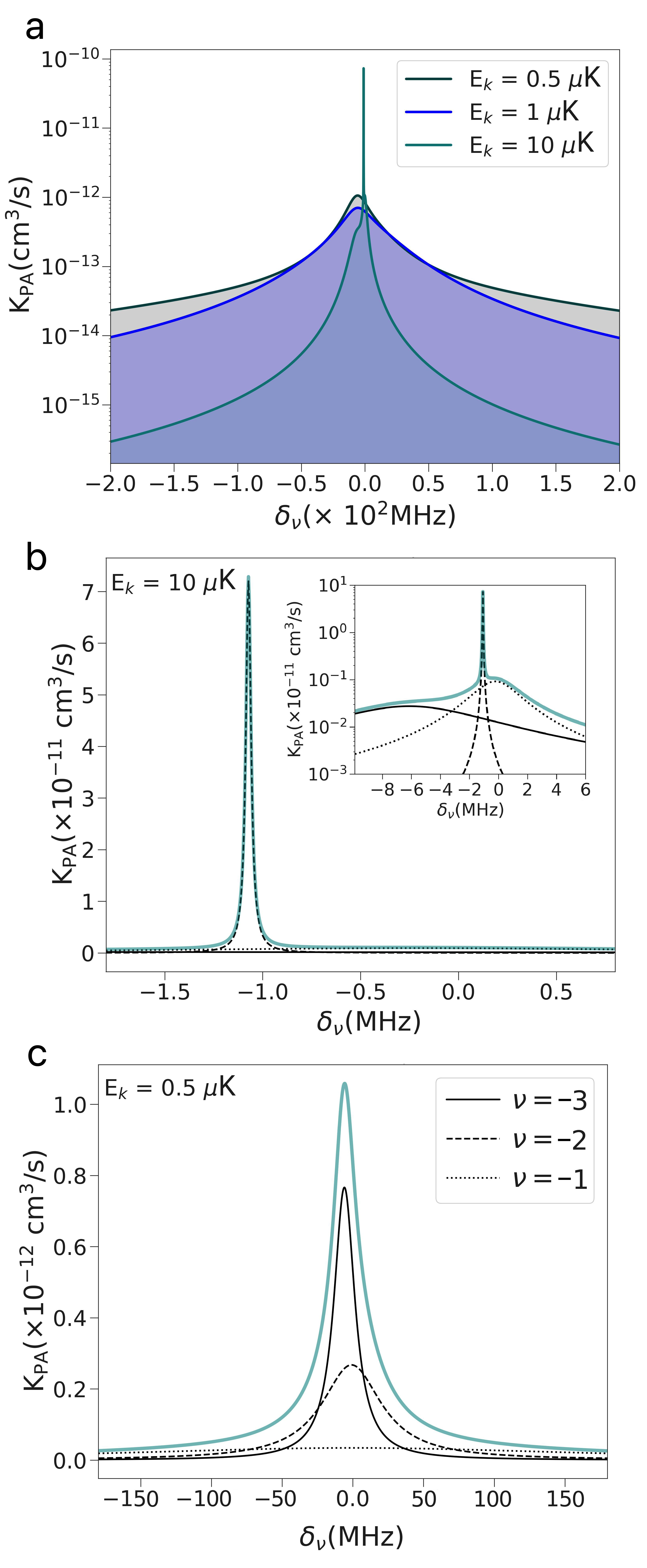}  
\caption{{\bf Photoassociation lineshape for LiRb+Rb$^*$(15 D$_{3/2}$) for different collision energies.} Panel (a) shows the PA lineshape for three collision energies. Panels (b) and (c) are the K$_{PA}$ constants for $E_{kin}$ = 10~$\mu$K and 500~nK, respectively, including the contributions from each vibrational state. The inset on panel (b) is a zoom-in of the PA peak. The color coding for panels (b) and (c) is the same: vibrational states are given in panel (c), and the total spectrum is in cyan. }
\label{fig:rate}
\end{figure}

We compute the two-photo PA lineshape profile for the LiRb-Rb(15 D$_{3/2}$), and the results are shown in figure \ref{fig:rate}. This figure displays the PA rate as a function of the two-photon detuning with respect to the least bound state associated with the LiRb-Rb(15 D$_{3/2}$) asymptote. The spectral overlap closely resembles the Lorentzian profile from equation \ref{eq3} due to the low energy values of the bound states relative to the kinetic energy. This causes all peaks to appear tightly centered around  $\delta_{\nu} = 0$. However, a closer examination of the spectrum, as shown in  in panel (b), reveals a more intricate structure where each bound state contributes uniquely. It is important to note that the peak intensity of each contribution varies with the collision energy
. The width of the PA spectra shows the expected inverse proportionality of the spectrum broadening with the temperature, as in figure \ref{fig:rate}.a .

\begin{figure}[htbp]
\centering
\includegraphics[width=0.44\textwidth]{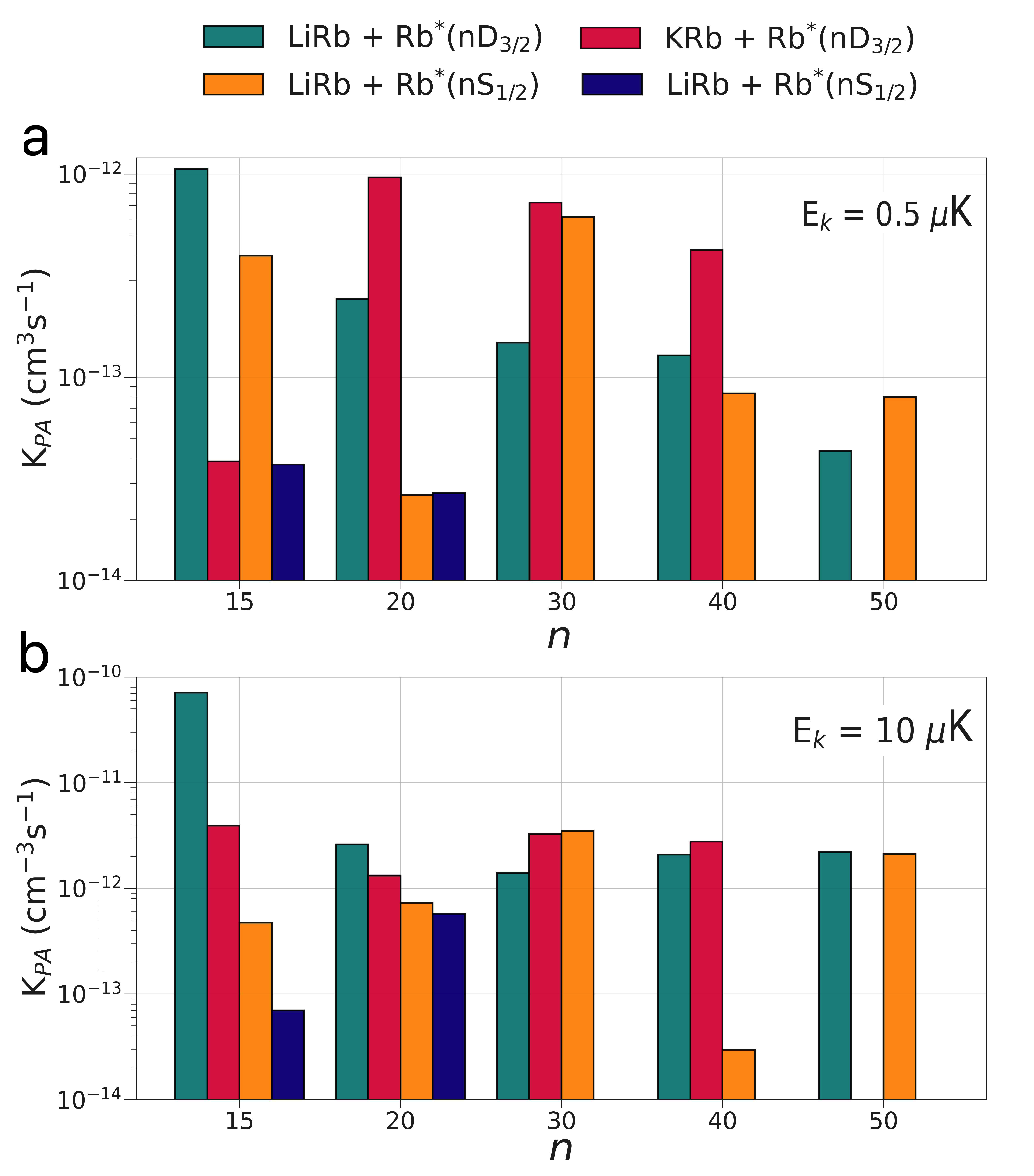}
\caption{{\bf Peak photoassociation rates constant for LiRb-Rb$^*$ and KRb-Rb$^*$.} Panels (a) and (b) displays the peak PA rate constants assuming a collision energy of 500~nK and 10~$\mu$K, respectively.}
\label{fig:rates}
\end{figure}

Finally, we report the PA rate peak for four different Rydberg-molecule combinations as a function of the principal quantum number, and the results are shown in figure \ref{fig:rates}. Independently of the species under consideration, the peak PA rate varies from 10$^{-13}$-10$^{-11}$cm$^{3}$s$^{-1}$. Clearly, systems involving Rydberg atoms in the D state outperform those with a Rydberg in the S state since the former shows larger C$_{6}$ coefficients. It seems that low principal quantum number states favor the PA rate toward the formation of long-range van der Waals Rydberg trimers, which arises from the dependence of the S-matrix elements on the bound state energies and the Franck-Condon factors (equation \ref{eq3}). The absence of bound states for sufficiently large 
$n$ values (Figure \ref{fig:depths_bounds}) is consistent with previous observations in the Rb$^{*}$+KRb system, in the context of inner-orbital long-range Rydberg trimers \cite{Ferez2017}.

\section{\label{conclusions}Conclusions}
We predict that long-range van der Waals Rydberg trimers exists in nature. The extension of these exotic trimers spans from 5 to 500~nm, depending on the atomic principal quantum number $n$ and orbital quantum number $L$. The size is comparable to other exotic Rydberg molecules such as trilobite and butterfly~\cite{Greene2000,Khuskivadze2002,Hamilton2002,Butterlfy,Trilobite,Panos2020}. Using the width of the PA absorption spectrum, We estimate the lifetime of the trimer states to range from 10$^{-9}$ to 10$^{-4}$~s, indicating that they could be synthesize in the lab via two photon PA.

To discuss the feasibility of observing these new exotic Rydberg molecules, first, we consider the stimulated absorption rate, which in our case is $\sim 10^{12}$s$^{-1}$. This number should be compared with PA rates measured for the formation of alkali-metal dimers from weakly excited (low $n$) alkali metal atoms. Experiments at  $100\, \mu$K  have measured stimulated rates of $ 1.5\times 10^8$ s$^{-1}$ for RbCs \cite{Kerman} and $3.5\times 10^7$ s$^{-1}$ for  LiRb \cite{Dutta}, using PA laser intensities of the order of $10^2$--$10^3$ W/cm$^{2}$. These examples involve species that interact also via van der Waals forces, but with values of $C_6$ that are several orders of magnitude smaller than our case. In addition, the PA process in these bi-akali systems occurs at relative short distances~\cite{Jones}, which makes the process more sensitive to the details of the initial scattering wavefunction. Therefore, the predicted stimulated absoprtion rate is indicative that the photoassociation between a Rydberg atom and a diatomic polar molecule into a long-range van der Waals Rydberg molecule is somewhat efficient. 

To observe long-range van der Waals Rydberg trimers, we propose to work on a regime where the molecular density is much smaller than the density of Rydberg atoms. We estimate the latter by requiring that at most one molecule resides within the Rydberg blockade radius $r_B$, which depends on the van der Waals interaction between atoms in identical Rydberg levels~\cite{Saffman2010}. Given a laser-dependent Rydberg excitation fraction $f<1$, the critical ground  atom density $\rho_B$ below which we can ignore blockade effects can be estimated  from $f\times \rho_B\times (4/3)\pi r_B^3=1$ \cite{Gallagher2005rydberg}. Assuming weak dressing $f\sim 10^{-2}$ and $r_B\sim 1\,\mu{\rm m}$ \cite{Singer_2005}, the relevant atom densities for this work $\rho_B\sim 10^{10}\,{\rm cm}^{-3}$ are thus compatible with magneto-optical trapping~\cite{SingerK,Tong}.

In comparison with traditional cold-molecule formation experiments that target high-density molecular ensembles ~\cite{Matsuda2020}, our Rydberg trimer predictions could be tested in a low-density regime where independent molecules are diluted in a background gas of alkali-metal atoms for atomic densities below the critical value beyond which Rydberg blockade effects become important~\cite{Balewski}. On the other hand, ultra-long-range van der Waals Rydberg trimers could be synthesized using optical tweezers. In this scenario, the molecule and atom are in adjacent tweezers and the atom in the next one, so the two-photon photoassociation profile is considered a function of the distance between the two tweezer traps. Thus, our work sets the foundations for a deeper understanding of the exotic properties of ultracold molecules in atomic Rydberg reservoirs.

\begin{table*}[ht]
\caption{Total PA rate constant peak for the Dimer+nD$_{3/2}$(Rb) at different temperatures. The rate peak is expressed in cm$^{3}$/s$^{-1}$}
\centering
\begin{tabular}{c c c c c c}
\hline
\hline
Molecule & $n$ & Vibrational energy (MHz) & $E_k = 0.5$ µK & $E_k = 1$ µK & $E_k = 10$ µK \\
\hline
\\
\multirow{3}{*} & \multirow{3}{*}{15} & $\nu_{-3}$ = 6.049   & \multirow{3}{*}{ $1.06 \times 10^{-12}$} & \multirow{3}{*}{$6.98 \times 10^{-13}$} & \multirow{3}{*}{$7.12 \times 10^{-11}$} \\
                      &                     & $\nu_{-2}$ = 8.783$\times$10$^{-1}$    &  &  &  \\

                      &                     & $\nu_{-1}$ =    3.916$\times$10$^{-3}$   \\
                      
                      \\
\multirow{3}{*}{} & \multirow{3}{*}{20} & $\nu_{-3}$ = 1.671 & \multirow{3}{*} {$2.43 \times 10^{-13}$} & \multirow{3}{*}{$4.79 \times 10^{-13}$ }& \multirow{3}{*}{$2.60 \times 10^{-12}$} \\
                      &                     & $\nu_{-2}$ = 2.347$\times$10$^{-1}$ &  &  &  \\
                      &                     & $\nu_{-1}$ = 1.492$\times$10$^{-3}$ &  &  &  \\
                      \\
\multirow{3}{*}{} {LiRb}& \multirow{2}{*}{30} & $\nu_{-2}$ = 1.341$\times$10$^{-1}$& \multirow{2}{*}{ $1.48 \times 10^{-13}$} & \multirow{2}{*}{$3.12 \times 10^{-13}$} & \multirow{2}{*}{$1.39 \times 10^{-12}$} \\
                      &                     & $\nu_{-1}$ = 2.404$\times$10$^{-3}$&  &  &  \\
                      \\
\multirow{2}{*}{} & \multirow{2}{*}{40} & $\nu_{-2}$ = 6.036$\times$10$^{-2}$& \multirow{2}{*} {$1.28 \times 10^{-13}$} & \multirow{2}{*}{$5.70 \times 10^{-13}$} & \multirow{2}{*}{$2.08 \times 10^{-12}$} \\
                      &                     & $\nu_{-1}$ = 2.301$\times$10$^{-3}$ &  &  &  \\
\\
\multirow{1}{*}{} & \multirow{1}{*}{50} & $\nu_{-1}$ = 1.061$\times$10$^{-2}$ & $4.33 \times 10^{-14}$ & $3.00 \times 10^{-12}$ & $2.21 \times 10^{-12}$ \\
                  \\
\hline
\hline
\multirow{1}{*}  & \multirow{1}{*}{15} & $\nu_{-1}$ = 2.741$\times$10$^{-2}$ & $3.85 \times 10^{-14}$ & $4.14 \times 10^{-14}$ & $3.92 \times 10^{-12}$ \\
\\
\multirow{1}{*}{KRb} & \multirow{1}{*}{20} & $\nu_{-1}$ = 2.808$\times$10$^{-3}$ & $9.62 \times 10^{-13}$ & $1.11 \times 10^{-13}$ & $1.32 \times 10^{-12}$ \\
\\     
\multirow{1}{*} & \multirow{1}{*}{30} & $\nu_{-1}$ = 2.201$\times$10$^{-3}$ & $7.23 \times 10^{-13}$ & $3.92 \times 10^{-13}$ & $2.25 \times 10^{-12}$ \\
\\
\multirow{1}{*}{} & \multirow{1}{*}{40} & $\nu_{-1}$ = 1.370$\times$10$^{-3}$& $4.24 \times 10^{-13}$ & $1.59 \times 10^{-13}$ & $2.77 \times 10^{-12}$ \\

\hline
\end{tabular}
\end{table*}

\begin{table*}[ht]
\caption{Total PA rate constant peak for the Dimer+nS$_{1/2}$(Rb) at different temperatures. The rate peak is expressed in cm$^{3}$/s$^{-1}$}
\centering
\begin{tabular}{c c c c c c}
\hline
\hline
Molecule & $n$ & Vibrational energy (MHz) & $E_k = 0.5$ µK & $E_k = 1$ µK & $E_k = 10$ µK \\
\hline
\\
\multirow{2}{*} & \multirow{2}{*}{15} & $\nu_{-2}$ = 1.257  & \multirow{2}{*}{ $3.96 \times 10^{-13}$} & \multirow{2}{*}{$3.94 \times 10^{-13}$} & \multirow{2}{*}{$4.72 \times 10^{-13}$} \\
                      &                     & $\nu_{-1}$ = 1.229$\times$10$^{-3}$    &  &  &  \\
                      \\
\multirow{1}{*}{} & \multirow{1}{*}{20} & $\nu_{-1}$ = 9.152$\times$10$^{-3}$& \multirow{1}{*} {$2.63 \times 10^{-14}$} & \multirow{1}{*}{$6.97 \times 10^{-14}$ }& \multirow{1}{*}{$7.38 \times 10^{-13}$} \\
\\
\multirow{1}{*}{} {LiRb}& \multirow{1}{*}{30} & $\nu_{-1}$ = 2.128$\times$10$^{-3}$& \multirow{1}{*}{ $6.15 \times 10^{-13}$} & \multirow{1}{*}{$3.25 \times 10^{-13}$} & \multirow{1}{*}{$3.47 \times 10^{-12}$} \\ \\
\multirow{1}{*}{} & \multirow{1}{*}{40} & $\nu_{1}$ = 1.396$\times$10$^{-3}$& \multirow{1}{*} {$8.31 \times 10^{-14}$} & \multirow{1}{*}{$4.83 \times 10^{-12}$} & \multirow{1}{*}{$2.95 \times 10^{-14}$} \\ \\
\multirow{1}{*}{} & \multirow{1}{*}{50} & $\nu_{-1}$ = 2.079$\times$10$^{-4}$ & $7.96 \times 10^{-14}$ & $1.35 \times 10^{-13}$ & $2.12 \times 10^{-12}$ \\
                  \\
\hline
\hline
\multirow{1}{*}  & \multirow{1}{*}{15} & $\nu_{-1}$ = 2.132$\times$10$^{-2}$ & $3.71 \times 10^{-14}$ & $7.30 \times 10^{-14}$ & $6.95 \times 10^{-14}$ \\
\\
\multirow{1}{*}{KRb} & \multirow{1}{*}{20} & $\nu_{-1}$ = 1.157$\times$10$^{-3}$ & $2.69\times 10^{-14}$ & $4.67 \times 10^{-14}$ & $5.74 \times 10^{-13}$ \\
\\

\hline
\end{tabular}
\end{table*}

\section{Aknowledgments}
M.L.C. and J.P.-R. acknowledge the support from the Stony Brook OVPR seed program and the the Simons Foundation. F.H. and V.O.-A. were supported by ANID through grants Fondecyt Regular 1221420 and Millennium Scientific Initiative ICN17\_012.
\appendix

\section{ Effective two-level system}
\label{appendixA}
The Hamiltonian that describes a three-level system as the one shown in Figure \ref{fig:scheme} is given by ($\hbar=1$)
\begin{equation}
\begin{aligned}
H &= \omega_e | e \rangle\langle e | + \omega_r | r \rangle\langle r | + \frac{\Omega_1}{2} \left[ |g \rangle\langle e| e^{i\omega_1 t} + |e \rangle\langle g| e^{-i\omega_1 t}  \right]\\
& \ \ \ \ + \frac{\Omega_2}{2} \left[ |e \rangle\langle r| e^{i\omega_2 t} + |r \rangle\langle e| e^{-i\omega_2 t}  \right],
\end{aligned}
\end{equation}
where $\omega_e$ ($\omega_r$) is the energy of the $|e \rangle$ ($|r \rangle$) state and the energy of $|g\rangle$ is set to zero. $\Omega_a$ and $\omega_a$ are the Rabi frequency and energy of the laser $a$, with $a=1,2$.

Using an unitary rotation frame transformation $U(t) = | g\rangle\langle g| + e^{i\omega_1 t} |e \rangle \langle e | + e^{i(\omega_1 + \omega_2)t} |r \rangle \langle r |$, the interaction Hamiltonian can be written as \cite{DanielBook}
\begin{equation}
\begin{aligned}
H_{I} &= - \Delta_1 | e \rangle\langle e | + \delta | r \rangle\langle r | \\
& \ \ + \frac{\Omega_1}{2} \left[ |g \rangle\langle e| + |e \rangle\langle g| \right] + \frac{\Omega_2}{2} \left[ |e \rangle\langle r|  + |r \rangle\langle e|   \right],
\end{aligned}
\end{equation}
where $\delta=\omega_r - (\omega_1 + \omega_2)$ and $\Delta_1 = \omega_1 - \omega_e$. The time evolution of the system is determined by the Schr\"odinger equation $i \partial_t |\Psi \rangle = H_{I} | \Psi \rangle$, where $|\Psi (t) \rangle = c_g(t) |g\rangle + c_e(t) |e\rangle + c_r(t) |r \rangle$ is the system state. In the rotating frame, $|e \rangle$ has fast oscillations and this state instantaneously tends to a steady state compared with the slow motion of the rest of the system, therefore we assume that $\dot{c}_e = 0$. Thus, the equations of motion are given by

\begin{equation}\label{eq:A1}
i\dot{c}_{g}(t) = \frac{\Omega_1^2}{4\Delta_1} c_g(t) + \frac{\Omega_1\Omega_2}{4\Delta_1} c_r(t),
\end{equation}

\begin{equation}\label{eq:A2}
i\dot{c}_r(t) = \delta c_r(t) + \frac{\Omega_2^2}{4\Delta_1} c_r(t) + \frac{\Omega_1\Omega_2}{4\Delta_1} c_g(t).
\end{equation}

Eqs. (\ref{eq:A1}) and (\ref{eq:A2}) are the same differential equation obtained from a two-level system with and effective Rabi frequency $\Omega_{\rm eff}= \frac{\Omega_1\Omega_2}{2\Delta_1}$ and an effective detuning

\begin{equation}
\delta_{\rm eff} = \delta + \frac{\Omega_2^2}{4\Delta_1} - \frac{\Omega_1^2}{4\Delta_1}.
\end{equation}

\newpage
\bibliography{apssamp}

\end{document}